\documentclass[pre,preprint,showpacs,preprintnumbers,amsmath,amssymb]{revtex4}
\usepackage{graphics,epstopdf,epsfig}
\begin {document}
\title{Global change in action due to trapping, how to derive it whatever the rate of variation of the dynamics.}
\author{Didier B\'enisti}
\email{didier.benisti@cea.fr} \author{Laurent Gremillet} 
\affiliation{ CEA, DAM, DIF F-91297 Arpajon, France.}
\date{\today}
\begin{abstract}
In this paper, we investigate the motion of a set of charged particles acted upon by a growing electrostatic wave, in the limit when the initial wave amplitude is vanishingly small and when all the particles have the same initial action, $I_0$. We show, both theoretically and numerically that, when all the particles have been trapped in the wave potential, the distribution in action exhibits a very sharp peak about the smallest action. Moreover, as the wave keeps growing, the most probable action tends towards a constant, $I_f$, which we estimate theoretically.  In particular, we show that $I_f$ may be calculated very accurately, when the particles' motion before trapping is far from adiabatic, by making use of a perturbation analysis in the wave amplitude. This fills a gap regarding the computation of the action change which, in the past, has only been addressed for slowly varying dynamics. 
Moreover, when the variations of the dynamics are fast enough, we show that the Fourier components of the particles' distribution function can be calculated by connecting estimates from our perturbation analysis with those obtained by assuming that all the particles have the same constant action, $I=I_f$. This result is used to compute theoretically the imaginary part of the electron susceptibility of an electrostatic wave in a plasma.  Moreover, using our formula for the electron susceptibility, we can extend the range in $\epsilon_a$ (the parameter that quantifies the slowness of the dynamics) for our perturbative estimate of $I_f-I_0$. This range can actually be pushed down to values of $\epsilon_a$ allowing the use of neo-adiabatic techniques to compute the jump in action. Hence, this paper shows that the action change due to trapping can be calculated theoretically, whatever the rate of variation of the dynamics, by connecting perturbative results with neo-adiabatic ones.
\end{abstract}
\maketitle
\section{Introduction}
As is well known~\cite{lenard}, for the nearly periodic and slowly varying dynamics of a Hamiltonian, $H(x,v,\varepsilon t)$,  the action $I$ defined as the area enclosed by a frozen orbit [i.e.,~an orbit of the Hamiltonian $H(x,v,\varepsilon t_0)$, where $t_0$ is a constant] is an adiabatic invariant. Nevertheless, it is also well known~\cite{neishtadt,cary} that the crossing of a separatrix (i.e.,~a frozen orbit that contains an unstable fixed point) breaks the adiabatic invariance, and the change in action, which quantifies the accuracy of the adiabatic approximation, has been extensively studied in the past due to its relevance to many fields of physics. To cite a few examples, action-variation calculations, and the adiabatic approximation itself, have been used in transport theory (see Refs.~\cite{leoncini,bazzani}~and references therein), celestial mechanics~(see for example Ref.~\cite{neishtadt2}), accelerator physics~\cite{cappi}, Bose-Einstein condensates (see  Ref.~\cite{itin} and references therein) and the nonlinear propagation of an electrostatic wave in a plasma (see for example Refs.~\cite{dodin,benisti07,benisti08,benisti09,benisti10,benisti12}) with an application to stimulated Raman scattering~\cite{benisti10b,benisti12b}. As regards the latter application, which motivated the present work, separatrix crossing occurs due to the trapping of electrons in the potential of an essentially growing electrostatic wave. This led us to focus, in this paper, on the motion of particles in an exponentially increasing potential and, for this important physics situation, we completely revisit the change in action, $\Delta I$, due to trapping. Indeed, we believe that the analysis we are presenting here significantly differs from the previous numerous publications on the subject in several respects.

First, we provide a theoretical estimate of $\Delta I$ for a non slowly varying dynamics, i.e., when the particles' motion before trapping is far from adiabatic. This fills a gap regarding the computation of the action change which, as far as we know, has always been performed within the framework of the neo-adiabatic theory~\cite{lenard,neishtadt,cary}, that is only useful for slowly varying dynamics. Consequently, as shown in Fig.~\ref{f6}, we are able to estimate $\Delta I$ when it is \textit{not} small compared to the initial action, which, to the best of our knowledge, has never been done before. At this point, one may wonder about the relevance of the action, $I$, for a non-slowly varying Hamiltonian dynamics, when the very notion of adiabatic invariance seems meaningless. Actually, for the adiabatic approximation to be valid, the period of a frozen orbit must be small compared to the typical time of variation of the Hamiltonian. Now, the period of a trapped orbit typically decreases as the square root of the potential amplitude so that, if this amplitude keeps growing, eventually, the variations in the action will become very small. Moreover, as we shall show it in this paper, and as illustrated in Fig.~\ref{f2}, provided that the growth rate is large enough then, by making use of a perturbation analysis in the potential amplitude, we are able to solve the equations of motion up to a time, $t_1$, when the amplitude is large enough for the action to remain nearly constant when $t>t_1$. Therefore, for large enough growth rates, perturbative results may be connected with adiabatic ones to provide an accurate solution of the equations of motion. In particular, as explained in Ref.~\cite{benistiDNC}, this procedure yields  the particles' distribution function at any time, for a non integrable dynamics when the classical methods of the neo-adiabatic theory do not apply! This shows the importance of computing the jump in action, $\Delta I$.

Second, we do not focus here on the microscopic description of the change in action for each particle, but, instead, we want to show how $\Delta I$ may be used to compute macroscopic, or averaged, quantities. To do so, we address the relevance of defining, at any time, one single action, $I^*(I_0)$, for a \textit{set} of particles having all the same initial action, $I_0$. Hence, by ``global change in action'', we mean here the change in $I^*(I_0)$, provided that this quantity is meaningful. Clearly, the concept of a ``global action'' for a set a particles, that we want to introduce here, would be exact if the distribution in action, $f(I,t)$, remained a Dirac distribution at any time, i.e., $f(I,t)=\delta (I-I^*)$. Therefore, we start by investigating if, and when, $f(I,t)$ has just one very sharp peak about a given action, $I^*$. Then, we study the convergence of $I^*$ towards an asymptotic value, $I_f$, as the wave amplitude keeps growing, and we define the global change in action as $\Delta I \equiv I_f-I_0$. 

Third, we test the usefulness of the concept of global action to compute one particular macroscopic quantity, $\chi_i$, which is the imaginary part of what may be viewed as a generalized electron susceptibility for  a plasma wave. 
Our definition for the generalized susceptibility, $\chi$, is given by Eq.~(\ref{chi}) of Section~\ref{III}, showing that $\chi$ is proportional to the ratio between the amplitude of the charge density and that of the wave. Then, Gauss law just translates into $1+\chi=0$, so that, necessarily, $\chi_i=0$. As discussed in several papers (e.g., Refs.~\cite{benisti07,benisti08,benisti09,benisti10,benisti12,benisti10b,benisti12b}), the latter equation may be used to derive such basic and important quantities as the nonlinear Landau damping rate of a plasma wave, and, more generally, describe very accurately the nonlinear propagation of such a wave. This is also true when the wave is laser-driven, so that, once the nonlinear variations of $\chi_i$ are known, one may address the nonlinear stage of stimulated Raman scattering, which has proven to be an issue for inertial confinement fusion~\cite{rosen}. We will not discuss here any of these points, that are way beyond the scope of this paper, and which will be the subject of a forthcoming article. Nevertheless, as an application of our results, we will show how to compute $\chi_i$ for an exponentially growing wave. This will actually let us estimate $\Delta I$ for a larger range in $\varepsilon_a$, the  parameter that quantifies the slowness of the dynamics, than by directly resorting to the distribution function, and, actually, down to values of $\varepsilon_a$ within the range of validity of the neo-adiabatic theory.  Hence, one important result of this paper is to show that is possible to provide an accurate estimate for the change in action due to trapping, whatever the rate of variation of the dynamics. This may be done by connecting results from a perturbation analysis, in the wave amplitude, with those obtained using the neo-adiabatic theory. 

Note that, in this paper, we focus on the asymptotic variation in action, $\Delta I \equiv I_f -I_0$, due to trapping so that, by ``global change in action'' we also mean the ``total action variation'' due to trapping. However, we also investigate the shift in action, $\delta I_S$, experienced by the particles once they have come close to the frozen separatrix. In particular, we investigate how $\delta I_S$ scales with $\varepsilon_a$, and compare this scaling with that of $\Delta I$. 

The previous points, which are the main results of our paper, are presented the following way. In Section~\ref{II}, we introduce the Hamiltonian dynamics that will be studied throughout the paper, and define what we will consider as the parameter, $\varepsilon_a$, that quantifies the slowness of the dynamics, and that usually is the small parameter of the neo-adiabatic theory (but which is not necessarily small here). When $\varepsilon_a$ is larger than unity, we show that a perturbation analysis in the wave amplitude may accurately describe the orbits in phase space up to the point when the amplitude is so large that the wave has trapped almost all the particles in its potential. Using this result, we can very easily predict the main features of the distribution in action, $f(I,t)$, when $f(I,t=0)=\delta(I-I_0)$. In particular, we can discuss when, and why, $f(I,t)$ should exhibit one single peak, about a given value $I^*$, which happens to be the smallest action. Moreover, the main results obtained theoretically for $f(I,t)$ when $\varepsilon_a \agt 1$ are numerically shown to remain valid when  $\varepsilon_a<1$, in particular as regards the fact that $f(I,t)$ eventually exhibits one single peak about $I^*$. The latter point is actually expected since, as we shall show it in Section~\ref{II}, it agrees with the predictions of the neo-adiabatic theory. The values of $I^*$ derived by making use of a perturbation analysis are compared to those calculated numerically, with an emphasis on the ability to correctly estimate the asymptotic value, $I_f$, reached by $I^*$ as the wave keeps growing. The comparison between the numerical and theoretical values of $I_f$ actually sets the limit, in $\varepsilon_a$, for the direct use of a perturbation analysis. The scaling of $\Delta I \equiv I_f-I_0$~with $\varepsilon_a$, in the limit $\varepsilon_a \rightarrow 0$, is also investigated numerically and compared to that obtained by making use of the neo-adiabatic theory.  Numerically, we also investigate the change in action, $\delta I_SÊ\equiv I^*-I_0$, when the particles are very close to the frozen separatrix, and the scaling of $\delta I_S$ with $\varepsilon_a$ is compared with that of $\Delta I$.

In Section~\ref{III}, we show that when $\varepsilon_a$ is large enough, the imaginary part, $\chi_i$, of the electron susceptibility of an electrostatic wave in a plasma may be accurately estimated by connecting perturbative results with those obtained by assuming that all the particles have the same constant action, $I=I_f$. Comparisons between the numerical and theoretical values of $\chi_i$ actually provide another diagnostic as regards the accuracy of our prediction for $I_f$. Moreover, since $\chi_i$ is more easily and more accurately computed than the distribution in action,  it can be used  to yield precise estimates of $I_f$ for a larger range in $\epsilon_a$ than in Section~\ref{II}.  In particular, we show that these perturbative estimates remain accurate down to values of $\varepsilon_a$ for which the scaling provided by the neo-adiabatic theory becomes valid. 

Finally, Section~\ref{iv}~summarizes and concludes this work. 

\section{The global change in action}
\label{II}
\subsection{The considered dynamics}
In the remainder of this paper, we will study the motion of particles in an exponentially growing sinusoidal potential, as given by the Hamiltonian
\begin{equation}
\label{eq1}
H_1=\frac{p^2}{2}-A_0 e^{\varepsilon \tau} \cos(x).
\end{equation}
Among all the physics problems that the Hamiltonian~(\ref{eq1}) could model, the present work was mainly motivated by issues regarding nonlinear wave-particle interactions in plasma physics~\cite{hdr}. More specifically, these issues are the very precise theoretical description of the nonlinear stage of the beam-plasma instability~\cite{bret}, and the derivation of the nonlinear Landau damping rate of a plasma wave in a very general situation that goes beyond the nearly adiabatic regime considered in Ref.~\cite{benisti07}, with application to backward Raman amplification~\cite{BRA}.  As will be shown in a forthcoming paper, in order to address these issues, it is essential to derive the global change in action due to trapping. To do so, we first define what we use as the small parameter, $\varepsilon_a$, that quantifies the slowness of the dynamics and, therefore, the accuracy of the adiabatic approximation. At first sight, it seems natural to use $\varepsilon_a=\varepsilon$, the wave growth rate, and to study how the change in action scales with $\varepsilon$, as done for example in Ref.~\cite{cary}. However, it is more accurate to use $\varepsilon_a=\varepsilon/p_0$, where $p_0$ is the initial value of $p$, since this represents the ratio between the period of a frozen orbit and the typical time of variation of the dynamics. Indeed, the period of an untrapped orbit, far from the separatrix, scales as $1/p_0$, while, for a nearly adiabatic motion, an orbit is trapped when $\sqrt{A}Ê\agt \pi p_0/4$~\cite{benisti07} (where $A\equiv A_0e^{\varepsilon \tau}$), and the period of this orbit then scales as $1/\sqrt{A}$. With this in mind, we now make the change of variables $t=\varepsilon \tau$, $v=p/\varepsilon$, and, in these new variables, the dynamics of $H_1$ is given by
\begin{equation}
\label{H2}
H=\frac{v^2}{2}-\Phi_0 e^{t} \cos(x),
\end{equation}
where $\Phi_0=A_0/\varepsilon^2$. The slowness of the dynamics defined by $H$ is quantified by $\varepsilon_a=1/v_0$, where $v_0$ is the initial value of $v$. Moreover, in the limit when $\Phi_0 \rightarrow 0$, which we will consider here, the initial action, $I_0$, is just $I_0=v_0$, since 
\begin{equation}
\label{action}
I=\frac{1}{2\pi}Ê\oint vdx,
\end{equation}
where the integral is over a frozen orbit, provided that this orbit is untrapped (for a trapped orbit, $I$ is defined as one half of the value given by Eq.~(\ref{action}), in order to avoid a jump in action only due to geometrical effects). We therefore conclude that $\varepsilon_a=1/I_0$. 

Note that if, for the dynamics of $H_1$, the change in action scales as $\varepsilon$, then the change in action for the dynamics defined by $H$ is just a constant, independent of $\varepsilon_a$.  Moreover, a change in action proportional to $\varepsilon \ln(\varepsilon)$ for the dynamics of $H_1$ would translate into an action change proportional to $\ln(\varepsilon_a)=-\ln(I_0)$, for the dynamics of $H$. 

In the remainder of this paper we will focus on the dynamics of $H$ in the limit $\Phi_0 \rightarrow 0$, and we will simply study the change in action as a function of $I_0$.

\subsection{The distribution in action}
\label{2.2}
Let us now investigate the distribution in action, $f(I)$, for a set of particles with the same initial action, $I_0$ (i.e., the same initial velocity), and whose positions are uniformly distributed between $0$ and $2\pi$. Since we consider the limit $\Phi_0 \rightarrow 0$, it is possible to describe the particles' motion, up to a certain time, by making use of a perturbation analysis in the amplitude of the potential, $\Phi \equiv \Phi_0e^t$. This has actually been done in Ref.~\cite{benisti07}, where it has been shown that $\varepsilon_p=\Phi/(1+I_0^2)$ may be chosen as the small parameter of the perturbative expansion, which should therefore provide accurate results when $\Phi \ll (1+I_0^2)$. Now, as regards the change in action, it mainly occurs when the orbit is close to the frozen sepratrix, i.e., when $\Phi \sim I_0^2$, so that $I$ should remain nearly constant once $\Phi \gg I_0^2$. Hence, when $I_0$ is sufficiently small compared to unity, is should be possible to use a perturbation analysis to derive the distribution in action up to the point when this distribution remains nearly stationary. Therefore, as explained in Ref.~\cite{benistiDNC}, by connecting perturbative results with adiabatic ones, it is possible to derive the particles' distribution function at any time! However, in this paper, we shall pursue another goal, which is the derivation of macroscopic quantities, such as moments or Fourier components of the distribution function, that are usually enough to address self-consistent physics problems, like the nonlinear propagation of waves in a plasma. Now, it is not necessary to go through the precise microscopic description of the distribution function to derive macroscopic quantities, and this would actually be very ineffective! In order to make this point more transparent, we start by investigating the main properties of the action distribution function, $f(I)$. 

\begin{figure}[!h]
\centerline{\includegraphics[width=18cm]{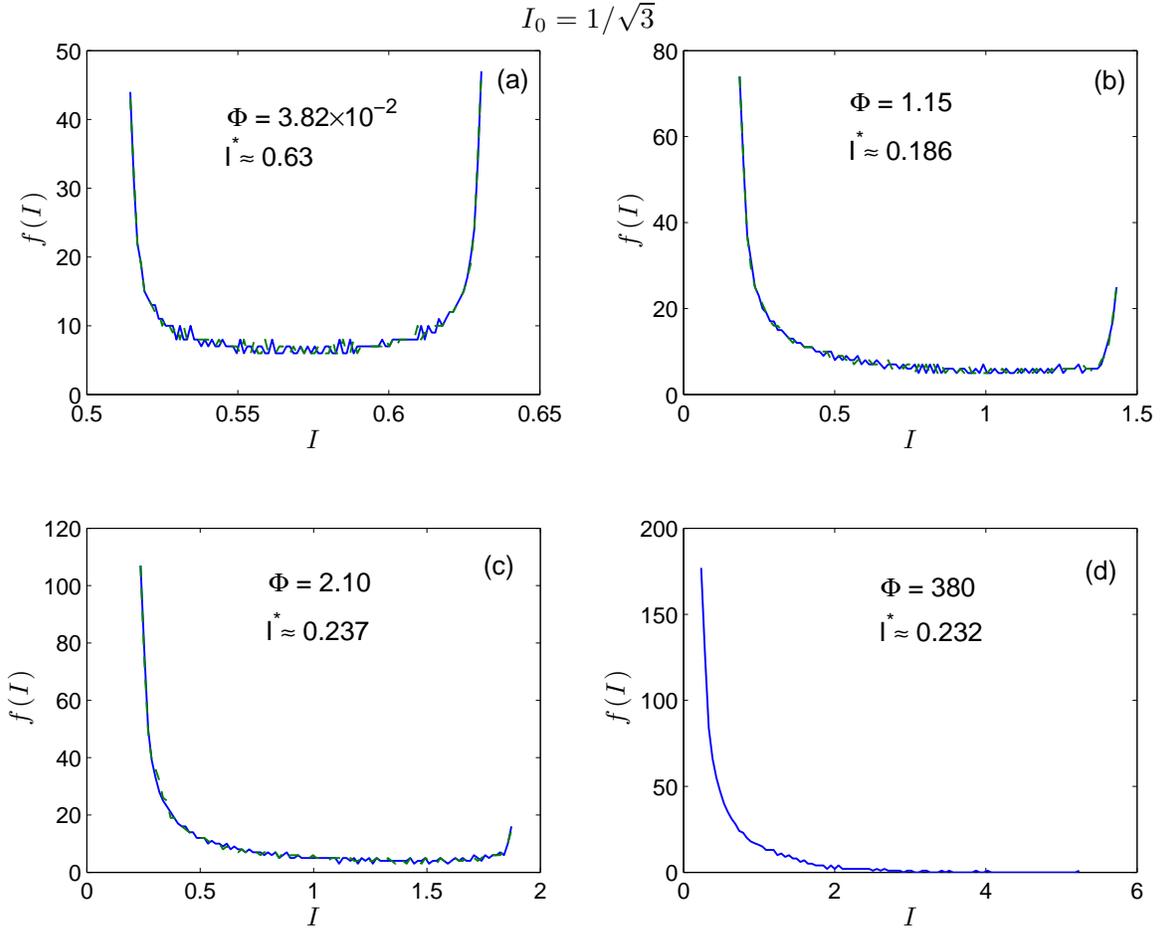}}
\caption{\label{f1}Distribution in action, $f(I)$, when $I_0=1/\sqrt{3}$ and for different values of $\Phi$. In panels (a), (b) and (c) is represented in blue solid line the distribution obtained numerically and in green dashed line that calculated perturbatively. In panel (d), only the numerical distribution function is plotted since a perturbation analysis is no longer valid for such a large amplitude. In each panel is indicated the value of the most probable action, $I^*$.}
\end{figure}

\begin{figure}[!h]
\centerline{\includegraphics[width=12cm]{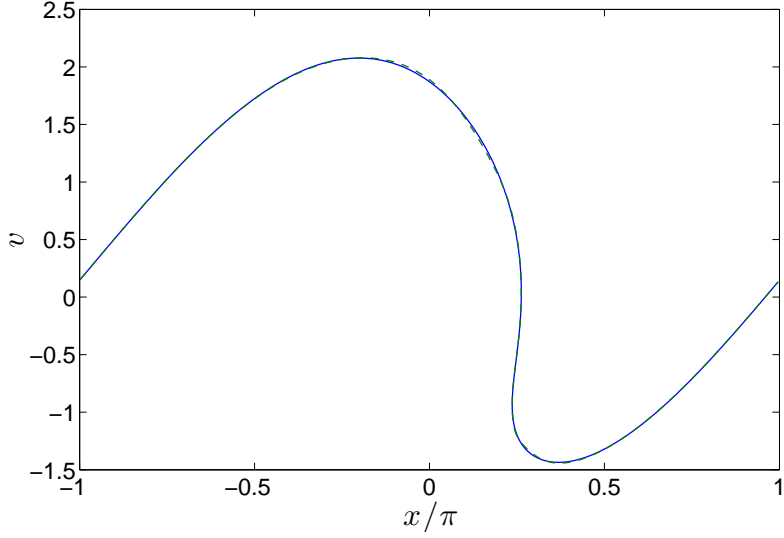}}
\caption{\label{f2}  Orbit in phase space calculated numerically (blue solid line) and perturbatively (green dashed line) when $I_0=1/\sqrt{3}$ and $\Phi=2.1$. For these parameters, more than 97\% of the particles are trapped in the potential, i.e., are such that $m<1$. }
\end{figure}

\begin{figure}[!h]
\centerline{\includegraphics[width=18cm]{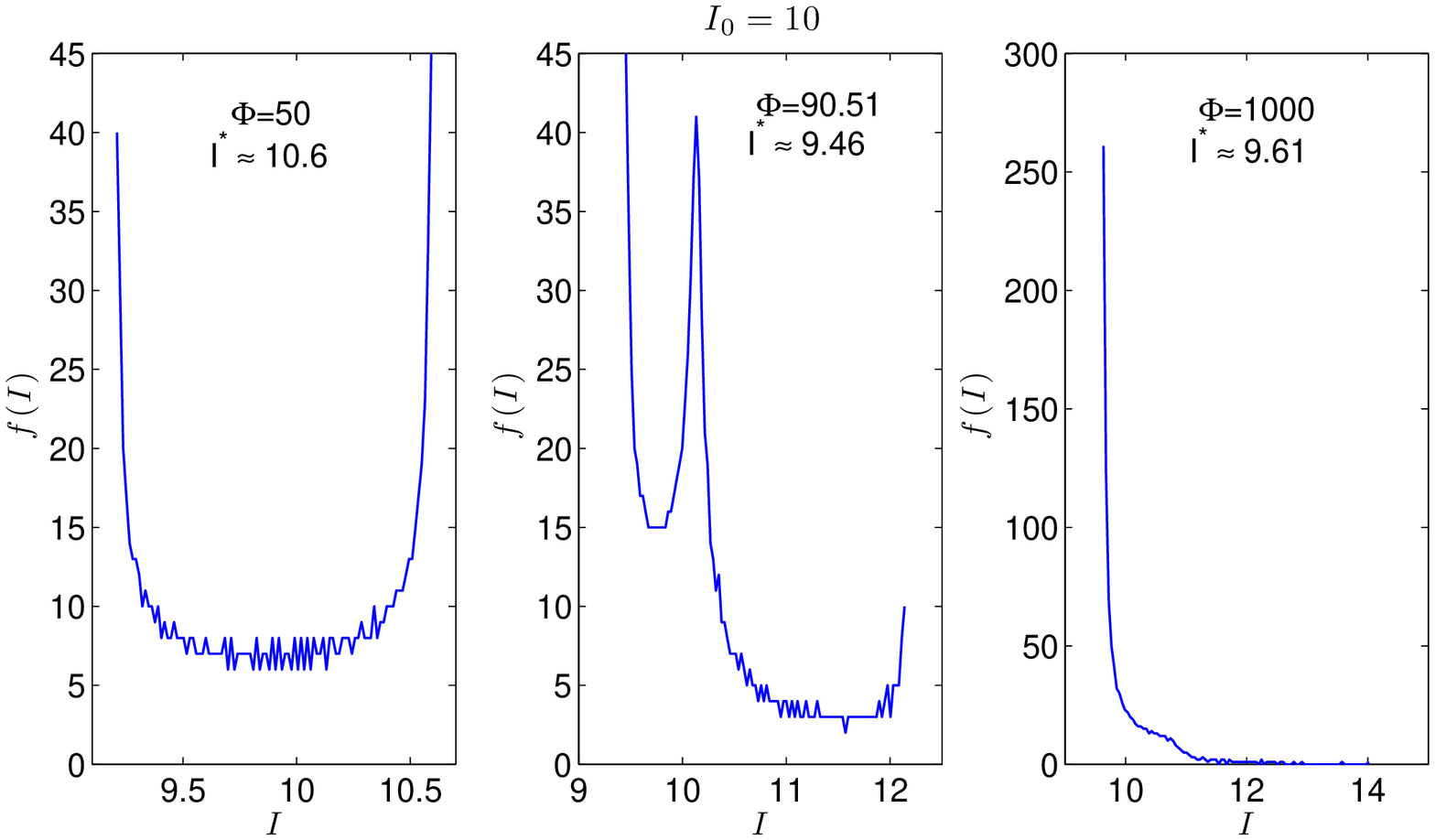}}
\caption{\label{f3} Distribution in action, $f(I)$, found numerically when $I_0=10$ and for various values of $\Phi$. }
\end{figure}

When the amplitude, $\Phi \equiv \Phi_0 e^t$, is so small that no particle is trapped in the potential, i.e., when
\begin{equation}
\label{m}
m \equiv \frac{H+\Phi}{2\Phi}
\end{equation}
is larger than unity for all particles, $f(I)$ exhibits two sharp peaks located at the minimum and maximum action. This may be seen in Fig.~\ref{f1}~(a)~comparing the perturbative results with those obtained numerically by directly solving the equations of motion with a symplectic leapfrog integrator~\cite{verlet}. Numerically, we choose $\Phi_0=10^{-8}$, we consider 1000 particles, and initialize them with the hypothesis that, when $\Phi_0 \rightarrow 0$, all particles have the same velocity (or, equivalently, the same action, $I_0$) and that their positions, $x_0$, are uniformly distributed between $0$ and $2\pi$ (see Ref.~\cite{benisti07}~for details). As may be seen in Fig.~\ref{f1}, when no particle is trapped, the perturbative analysis (led, here, up to to order 12~\cite{explication}), is very accurate, and lets us understand very easily why the distribution in action has two sharp peaks. Indeed, from a perturbative expansion, and for each amplitude, $\Phi$, one can express the action of any particle as a function of its initial position, $x_0$, and of its initial velocity. Since we consider here the situation when all particles have the same initial velocity and when $x_0$ is uniformly distributed between $0$ and $2\pi$, we conclude that the distribution function in action, $f(I)$, should just be proportional to $(\partial  I/\partial x_0)^{-1}$. Therefore, if the function $I(x_0)$ has some extrema, $f(I)$ should be very peaked about each of these extrema. Now, we find that the function $I(x_0)$ calculated perturbatively has just one maximum and one minimum, which explains why $f(I)$ has exactly two peaks about the minimum and maximum action. These two peaks nearly have the same amplitude, although that located at the maximum action is a bit higher, so that, for the corresponding values of  $\Phi$, the most probable action, $I^*$, is the maximum one. 

As the amplitude, $\Phi$, keeps on increasing, more and more particles are trapped in the potential, i.e., are such that $m$, as defined by Eq.~(\ref{m}), is less that unity. Even in the situation when a large fraction of particles are trapped, the perturbative analysis remains very accurate to describe the distribution in action, as may be seen in Fig.~\ref{f1}~(c), and gives a very good approximation the orbit in phase space, as shown in Fig.~\ref{f2}. Hence, even when a large amount of particles are trapped, by using the same argument as before, we conclude that $f(I)$ should have two sharp peaks about the minimum and maximum actions, which is indeed the case as illustrated in Fig.~\ref{f1}~(b). However, these two peaks do not have the same height because, now, the minimum action is for trapped particles while the maximum action is for the untrapped ones, so that the relative amplitude of the two peaks is just proportional to the relative abundance of these two distinct types of particles. Hence, as long as perturbative results are accurate, we can prove that, as $\Phi$ keeps on increasing and more and more particles are getting trapped, the peak in $f(I)$ located at the minimum action becomes more and more prominent while that located at the maximum action tends to vanish. Now, when $I_0 \alt 1$,  the perturbative analysis can be led up to the point when nearly all the particles are trapped in the potential so that $f(I)$ exhibits one single peak at the minimum action, $I_{\min}$ [see Fig.~\ref{f1}~(c)]. At this point, the particles with $I=I_{\min}$ are deeply trapped, i.e., they are far away from the frozen separatrix, so that, as $\Phi$ keeps on increasing their action does not vary much, and $f(I)$ keeps one single peak at $I=I_{\min}$, as shown in Fig.~\ref{f1}~(d). Note that, in Fig.~\ref{f1}~(c)~for $\Phi=2.1$, $I_{\min}\approx0.237$, while in Fig.~\ref{f1}~(d)~for $\Phi=380$, $I_{\min}\approx0.232$.   Therefore, when $I_0\alt 1$, we are able to prove that, eventually, once all the particles have been trapped in the potential, $f(I)$ has one single sharp peak at the value, $I=I_{\min}$, that remains nearly constant. 

For larger values of $I_0$, we resort to numerical simulations in order to study the variations of $f(I)$, as $\Phi$ increases.  Figs.~\ref{f3}~(a)-(c) show the evolution of $f(I)$ with $\Phi$ when $I_0=10$. When $\Phi$ is so small that most particles are untrapped and perturbative results are accurate then, for the same reason as before, $f(I)$ has two sharp peaks about the minimum and maximum actions. For intermediate values of $\Phi$, when most particles are trapped and the perturbative expansion is no longer valid then, as may be seen in Fig.~\ref{f3}~(b), a new peak in $f(I)$ may appear for an action slightly larger than the minimum one, a feature which is found numerically and that cannot be explained with the theoretical arguments used when $I_0 \alt 1$. Nevertheless, as $\Phi$ keeps on increasing, and for all the cases we investigated numerically, we found that, eventually, $f(I)$ exhibited only one sharp peak and that the most probable action, $I^*$, was also the minimum one. 

The latter result is actually expected for large values of $I_0$ from neo-adiabatic theory. Indeed, from Eq.~(83) of Ref.~\cite{cary}, we conclude that, as $\Phi$ increases, the action of any particle should converge to the value $I_{\infty}$ given by, 
\begin{equation}
\label{Ii}
I_{\infty}=I_0-(2/\pi) \ln \left\vert 2\pi\sin\left(h_0/I_0\right)\right\vert,
\end{equation}
where $h_0$ is the value of $(H-\Phi)$ when the particle crosses the line $x=\pi$ (modulo $2\pi$) for the last time before being trapped in the potential. The values of $h_0$ may be found by using Eqs.~(2.12), (2.17) and (2.19) of Ref.~\cite{nonlinearity}. These equations let us conclude that, for the case considered in this paper, with positions uniformly distributed between 0 and $2\pi$ in the limit $\Phi \rightarrow 0$, $h_0=-\pi I_0u$, where $u$ is uniformly distributed between zero and unity. Plugging this expression for $h_0$ into Eq.~(\ref{Ii}), we find,
\begin{equation}
\label{Ii2}
I_{\infty}=I_0-(2/\pi) \ln \left\vert 2\pi\sin\left(\pi u\right)\right\vert,
\end{equation}
which clearly shows that $I_{\infty}$ has only one extremum, which actually is an absolute minimum. Since $u$ is uniformly distributed, we conclude that neo-adiabatic theory does predict that $f(I)$ should eventually exhibit only one peak at the minimum action. Moreover, Eq.~(\ref{Ii2}) provides an explicit simple expression for the asymptotic value, $I_f$, of the most probable action, $I_f=I_0-(2/\pi)\ln(2)$. We therefore conclude that, for large enough values of $I_0$, the global change in action, $\Delta I\equiv I_0-I_f$, as predicted by the neo-adiabatic theory, is 
\begin{equation}
\label{Di}
\Delta I=-(2/\pi)\ln(2)\approx -0.441,
\end{equation}
a result we shall now check numerically in the Subsection~\ref{IIC}. Note that $\Delta I$ given by Eq.~{(\ref{Di}) is independent of $I_0$, meaning that, if one used Hamiltonian $H_1$ given by Eq.~(\ref{eq1}), one would find that the action change would scale as $\epsilon$ in the limit $\epsilon\rightarrow 0$.

\subsection{Asymptotic value of the most probable action.}
\label{IIC}
The typical evolution of the most probable action, $I^*$, as a function of the wave amplitude, $\Phi$, is plotted in Fig.~\ref{f4}~when $I_0=1$. For small values of $\Phi$, $I^*>I_0$, and it slightly increases with $\Phi$ because, for such small amplitudes, when most particles are untrapped then, as seen in Subsection~\ref{2.2}, $I^*$ is the maximum action.  
\begin{figure}[!h]
\centerline{\includegraphics[width=12cm]{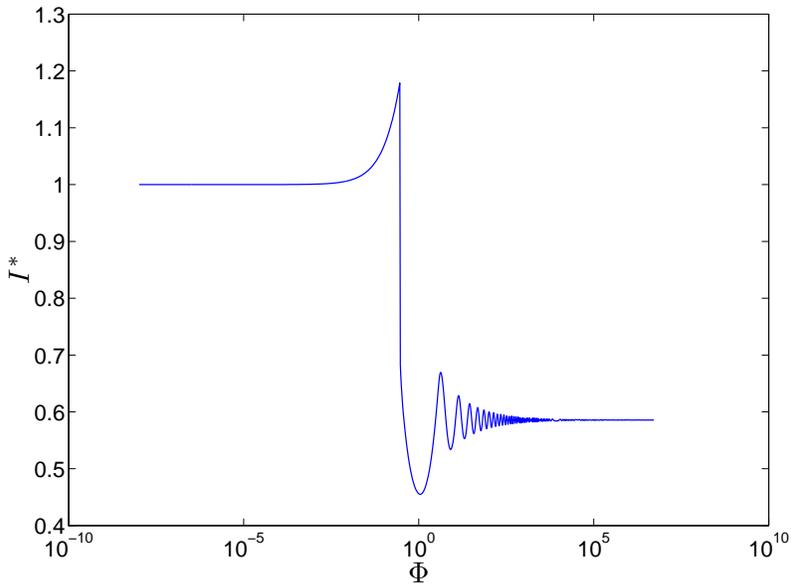}}
\caption{\label{f4} Evolution of the most probable action, $I^*$, with $\Phi$ when $I_0=1$. }
\end{figure}

A sudden jump, $\delta I$, occurs in $I^*$ when most particles are trapped so that $I^*$ no longer is the maximum action but the minimum one. Therefore, $\delta I$ is the difference between the maximum and minimum action, hence the spread in action,  when the particles are close to the frozen separatrix.  As shown in Fig.~\ref{f5}, $\delta I$ scales as $\ln(I_0)$, when $I_0\agt 10$ (which means that, for the Hamiltonian $H_1$ given by Eq.~(\ref{eq1}), the spread in action would scale as $\varepsilon \ln(\varepsilon)$). Fig.~\ref{f5} also shows that the change in action, $\delta I_S$, for particles close to the separatrix, either trapped or untrapped, scales as $\ln(I_0)$. Hence, although studying $\delta I_S$ in detail is beyond the scope of this paper, it is important to note that the total change in action, $\Delta I$, is not the maximum action shift experienced by the particles as $\Phi$ increases. Actually,  $\Delta I$, as predicted by Eq.~(\ref{Di}),  does not even scale with $I_0$ as the maximum action shift.

\begin{figure}[!h]
\centerline{\includegraphics[width=12cm]{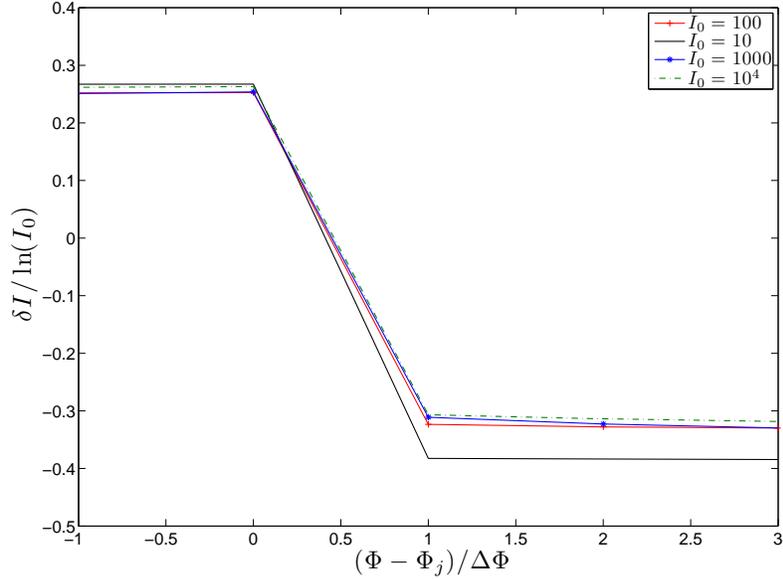}}
\caption{\label{f5} Jump, $\delta I$, in the most probable action divided by $\ln(I_0)$ when $I_0=10$ (black solid line), $I_0=100$ (red solid line with pluses), $I_0=1000$ (starred blue solid line) and $I_0=10^4$ (green dashed line). The amplitude, $\Phi$, has been centred about the value, $\Phi_j$, where the jump occurs, and rescaled so that, for our numerical data, we see a jump in $I^*$ when the amplitude changes by 1.  }
\end{figure} 

\begin{figure}[!h]
\centerline{\includegraphics[width=12cm]{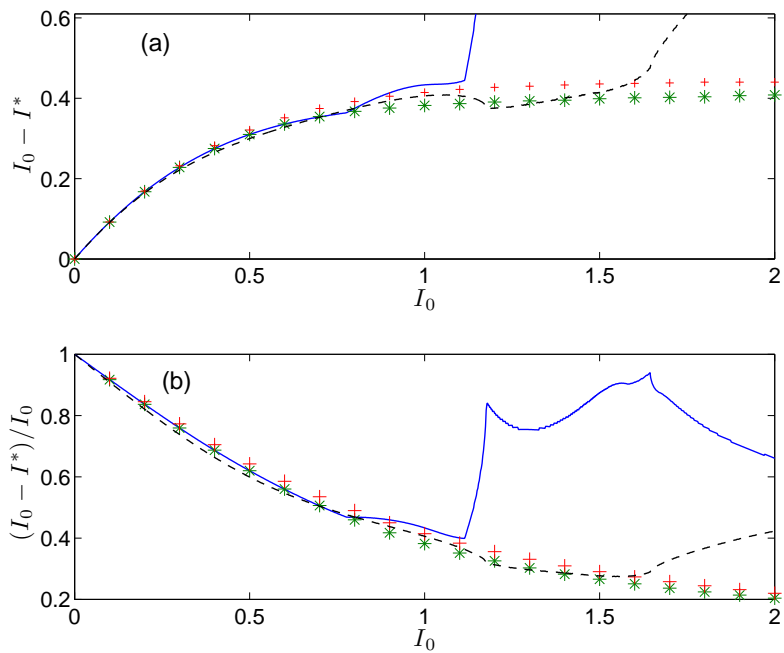}}
\caption{\label{f6} Panel (a), change in action, $I_0-I^*$, as a function of $I_0$. The blue solid line plots this change when $I^*$ is derived from the perturbative distribution function calculated when $\chi_i$ reaches its first maximum. The green stars also correspond to values of $I^*$ at the first maximum of $\chi_i$, but they are deduced from the distribution function calculated numerically. The black dashed line also refers to $I^*$ at the fist maximum of $\chi_i$, and it is evaluated by making use of Eq.~(\ref{C2}) of Section~\ref{III}. The red pluses plot $(I_0-I_f)$ as estimated numerically. In panel (b) is plotted the relative change in action, $(I_0-I^*)/I_0$, with the same conventions as in panel (a).}
\end{figure} 

\begin{figure}[!h]
\centerline{\includegraphics[width=12cm]{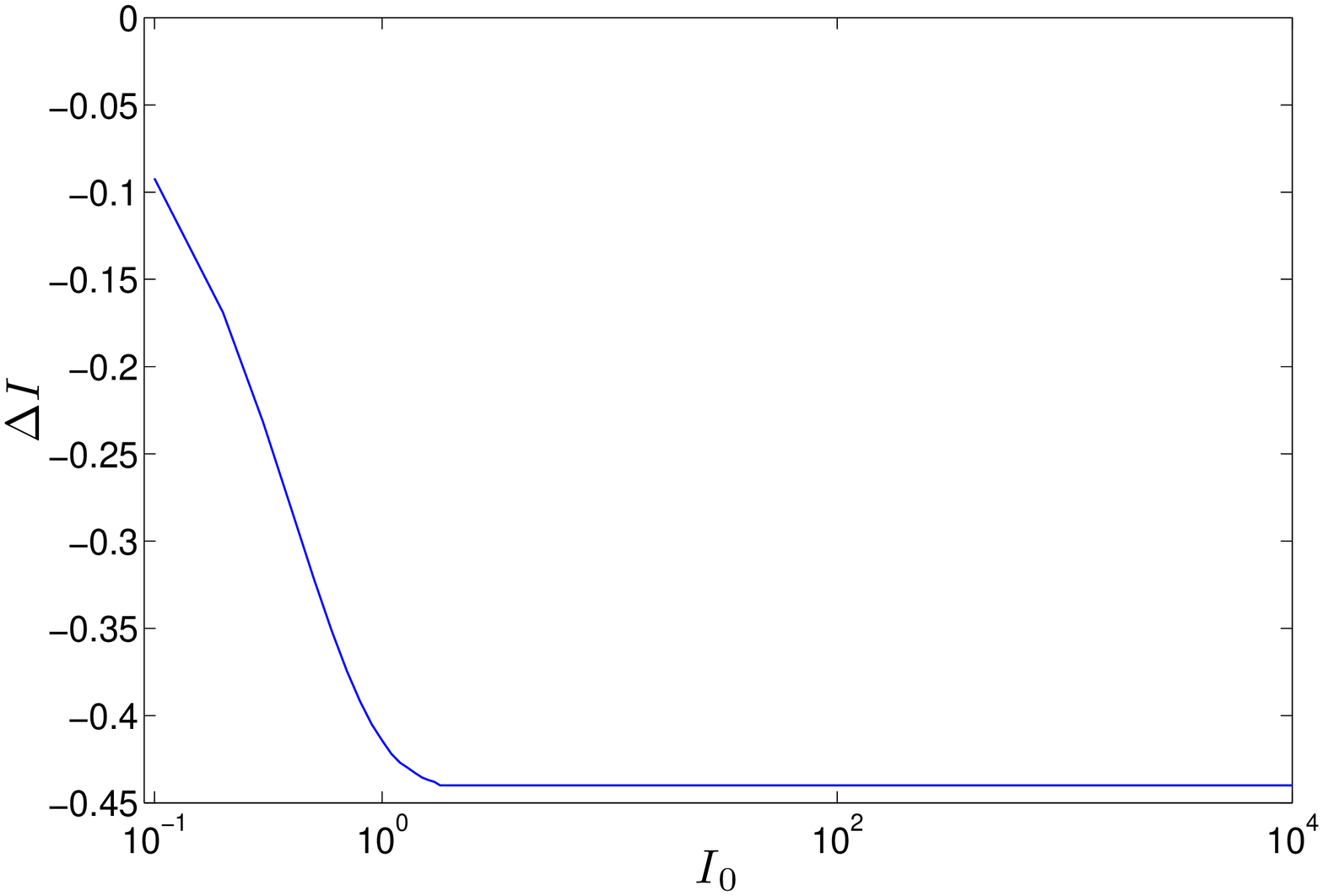}}
\caption{\label{f7} Global change in action, $\Delta I = I_f-I_0$, as a function of $I_0$}
\end{figure} 

After the jump, $I^*$ exhibits some oscillations of smaller and smaller amplitude and therefore seems to converge towards an asymptotic value, $I_f$, as the amplitude of the potential keeps on increasing. We now investigate how accurately $I_f$ may be estimated theoretically by making use of a perturbative expansion. When doing so, we cannot take the limit $\Phi \rightarrow \infty$, because the perturbative expansion is limited  to a finite range of amplitudes.  Therefore, we identify $I_f$ with the value of the most probable action, $I^*$, at a given amplitude, $\Phi_M$, large enough for $I^*(\Phi)$ to remain nearly constant when $\Phi>\Phi_M$ and, yet, small enough to remain within the range on validity of the perturbative analysis. Since we want to apply our results on the action change to the computation of $\chi_i$, the imaginary part of the electron susceptibility defined by Eq.~(\ref{chi}) of Section~\ref{III}, we choose $\Phi_M$ as the amplitude when $\chi_i$ reaches its first maximum. Indeed, as we shall show it in Section~\ref{III}, after reaching its first maximum, $\chi_i$ oscillates with $\Phi$ in a very regular fashion, thus reflecting the nearly adiabatic motion of trapped particles, i.e., the near constancy of their action. As may be seen in Fig.~\ref{f6}, when $I_0\alt 1$, the perturbative value of $I^*$ at the amplitude when $\chi_i$ reaches its first maximum is in excellent agreement with the numerical one, and it does indeed provide a very good estimate of $I_f$, which is only underestimated by about 5\%. Hence, as clearly shown in Fig.~\ref{f6}, we are indeed able to precisely calculate the global change in action due trapping, $\Delta I \equiv I_f-I_0$, even when this change is of the order of the initial action. 

As may be already guessed from Fig.~\ref{f6}~(a), and is obvious in Fig.~\ref{f7}, the global change in action, $\Delta I =I_f-I_0$, converges towards a constant as $I_0 \rightarrow \infty$. Numerically, this constant is found to be very close to $-0.44$ (see also Figs.~\ref{f9} and~\ref{f10}), which is in excellent agreement with the prediction of Eq.~(\ref{Di}) from neo-adiabatic theory. Note also that the convergence of $\Delta I$ towards a constant occurs quite rapidly since we numerically estimate that, when $I_0=1$, $\Delta I \approx -0.414$ [which departs by less than 10\% from the asymptotic value of Eq.~(\ref{Di})], while, when $I_0=1.6$, $\Delta I \approx -0.437$ [which departs by less than 1\% from the asymptotic value of Eq.~(\ref{Di})]

\section{Application to the derivation of the imaginary part of the electron susceptibility for an electrostatic wave in a plasma}
\label{III}
In Section~\ref{II}, we showed that we could provide a theoretical estimate for the global change in action, $\Delta I$, when it is not small compared to the initial action, i.e., when the classical techniques of the neo-adiabatic theory do not apply. However, to do so, we had to use a perturbation analysis up to order 12, and the corresponding formulas are pages long so that, although it is important to provide theoretical results, one may wonder about the practical interest of such theoretical developments\dots~Moreover, one may also wonder about the physics relevance of the most probable action, $I^*$, we use to define $\Delta I$. We shall now answer these questions by showing that, indeed, $I^*$ is useful to compute the imaginary part, $\chi_i$, of the electron susceptibility for a plasma wave, and that $\chi_i$ will actually provide a  very fine diagnostic for our prediction of $\Delta I$. Moreover, we will show that, by using our perturbative results for $\chi_i$, which are actually published in Ref.~\cite{benisti07}~and are much more simple than those giving the orbit in phase space, we are able to calculate $\Delta I$ for a larger range in $I_0$ than in Section~\ref{II}. 

\subsection{The electron susceptibility}
The sinusoidal potential used in Hamiltonian $H$, Eq.~(\ref{H2}), may be viewed as the potential of a sinusoidal electric field, 
\begin{eqnarray}
\label{E}
E&=&-i\frac{\Phi}{2}e^{ix}+c.c. \\
&\equiv &E_0e^{ix}+c.c.,
\end{eqnarray}
where $c.c.$ stands for the complex conjugate. This field induces, for example in a plasma, the charge density,

\begin{equation}
\label{rho}
\rho =\rho_0 e^{ix}+c.c.,
\end{equation}
and we introduce 
\begin{equation}
\label{chi}
\chi \equiv \frac{i\rho_0}{\varepsilon_0 E_0},
\end{equation}
so that, since $E_0$ only depends on time, Gauss law just reads,
\begin{equation}
\label{gauss}
 1+\chi=0.
\end{equation}
Note that the electron susceptibility is usually defined in Fourier space, while we define here $\chi$ is the direct space, because the use of Fourier representation is of little help to address the nonlinear regime of wave-particle interaction we focus on in this paper. The imaginary part of Eq.~(\ref{gauss}) simply yields $\chi_i=0$, and, as shown in previous papers (see Ref.~\cite{benisti12} and references therein), the resolution of this equation, with $\chi_i$ derived for an exponentially growing wave, could provide values of such complex quantities as Raman reflectivity in a plasma, in the nonlinear kinetic regime once Landau damping has nearly vanished, and in a three-dimensional geometry. However, previous results were only for a nearly adiabatic situation, with a smooth distribution in the initial particles' velocity, so that phase mixing was effective enough to render negligible the contribution to $\chi_i$ from trapped particles (see Ref.~\cite{benisti07,yampo}). We now want to calculate this contribution very precisely, which lets us choose an initial condition with the same initial velocity for all particles, so that phase mixing cannot occur. 

Note now that, form Eq.~(\ref{rho}), since $\rho_0$ only depends on time, 
\begin{eqnarray}
\rho_0&=&\frac{1}{2\pi}Ê\int_{-\pi}^{\pi} \rho e^{-ix} dx \\
&=& \frac{1}{2\pi}Ê\int_{-\pi}^{\pi} \int_{-\infty}^{+\infty}F(x,v,t) e^{-ix} dxdv \\
&\equiv& \langle e^{-ix} \rangle,
\end{eqnarray}
where $F(x,v,t)$ is the particles' distribution function, and where $\langle . \rangle$ stands for the statistical averaging over all particles. Since, from Eq.~(\ref{E}), $E_0$ is purely imaginary, we conclude that $\chi_i$ is proportional to $\langle \sin(x) \rangle$, the quantity we focus on in the remainder of this paper. Note that, for a discrete set of particles, as considered in numerical simulations,
\begin{equation}
\langle \sin(x) \rangle =\frac{1}{N}Ê\sum_{i=1}^N \sin(x_i),
\end{equation}
where $N$ is the total number of particles and $x_i$ is the position of the $i^{th}$ particle.

\subsection{Use of the global change in action to compute $\chi_i$}

\subsubsection{Theoretical estimate of $\chi_i$ and comparisons with numerical results}
In this subsection we show that, when $I_0$ is small enough, it is possible to compute $\chi_i$ (or $\langle \sin(x) \rangle$) by connecting perturbative estimates with adiabatic ones. We, therefore, make full use of the results of Section~\ref{II} showing that a perturbative expansion may be accurate up to amplitudes beyond which the action remains nearly constant. The perturbative value of $\langle \sin(x) \rangle$, up to order 11~\cite{explication}, may be found in Ref.~\cite{benisti07}. We now assume that the perturbative estimate of $\langle \sin(x) \rangle$ remains accurate up to large enough values of $\Phi$ so as to correctly predict the first maximum of $\langle \sin(x) \rangle$, which we denote by $S_M$, and the corresponding value, $\Phi_M$, of the wave amplitude (as shall be seen in the next subsection, this is indeed the case when $I_0\alt 1.6$~\cite{explication2}). For values of $\Phi$ larger than $\Phi_M$ we shift to action-angle variables $(\theta,I)$ in order to compute $\langle \sin(x) \rangle$, namely, we introduce $\tilde{f}(\theta,I,t)=F(x,v,t)$ the action-angle distribution function, to find,
\begin{equation}
\label{0}
\langle \sin(x) \rangle = \frac{1}{2\pi}Ê\int_0^{2\pi} \int_{0}^{+\infty} \sin[x(\theta,I)] \tilde{f}(\theta,I,t) d\theta dI
\end{equation}
(note that the change of variables $(x,v) \rightarrow (\theta,I)$ is canonical so that its Jacobian is unity). We now assume that, by the time $\Phi$ reaches the value $\Phi_M$, all the particles have been trapped in the potential. Then, using the formulas of $\theta$ and $I$ for trapped particles (see Ref.~\cite{benisti07}), we find
\begin{eqnarray}
\sin(x)&=&2\sin(x/2)\cos(x/2) \\
&=& 2 \sqrt{m}~\text{sn}\left[Ê\left. \frac{2 K \theta}{\pi}Ê\right \vert m\right]\times \text{dn}\left[Ê\left. \frac{2 K \theta}{\pi}\right\vert mÊ\right],
\end{eqnarray}
where $Ê\text{sn}(u\vert m)$ and $\text{dn}(u \vert m)$ are Jacobian elliptic functions and $K\equiv K(m)$ is the complete elliptic integral of first kind~\cite{abra}. Using the Fourier representation of elliptic fonctions \cite{abra}, we find,
\begin{equation}
\sin(x) =2 \left \{ \frac{2\pi}{K} \sum_{n=0}^{+\infty} \frac{q^{n+1/2}}{1-q^{2n+1}} \sin[(2n+1)\theta] \right \} \times 
\left \{\frac{\pi}{2K}+\frac{2\pi}{K} \sum_{n=1}^{+\infty} \frac{q^{n}}{1+q^{2n}} \cos[2n\theta] \right \},
\end{equation}
where $q \equiv \exp[-\pi K(1-m)/K(m)]$. Note that $q<1$ and that it rapidly decreases with $m$. Hence, if all the particles are deeply trapped when $\Phi>\Phi_M$, i.e., are such that $m$ is significantly less than unity, then $ \sin(x)$ may be approximated by its first Fourier coefficient in $\theta$, namely,
\begin{equation}
\label{e0}
\sin(x) \approx \frac{2Ê\pi^2}{K^2}Ê\frac{\sqrt{q}}{1-q}Ê\sin(\theta).
\end{equation}
Now, in Section~\ref{II} we saw that, once all the particles have been trapped, the distribution in $I$ exhibits one sharp peak and that, once $\langle \sin(x) \rangle$ has reached its first maximum, the most probable action is quite close to its asymptotic value, $I_f$. Consequently, we may approximate $\langle \sin(x) \rangle$ by assuming that, when $\Phi>\Phi_M$, all particles have the same constant action, $I=I_f$. This allows us to relate the angle $\theta$ of each particle to the value, $\theta_M$, reached at $t=t_M$ when $\Phi=\Phi_M$, by the same formula,
\begin{equation}
\label{ee0}
\theta(t)=\theta_M +\int_{t_M}^t \omega_0(I_f)dt',
\end{equation}
with 
\begin{equation}
\label{e1}
\omega_0(I_f)=\frac{\pi \sqrt{\Phi}}{2K[m(I_f)]},
\end{equation}
and $m(I_f)$ is such that,
\begin{equation}
\label{e2}
\frac{4 \sqrt{\Phi}}{\pi} \left\{E[m(I_f)]+Ê[m(I_f)-1]K[m(I_f)]\right\}=I_f,
\end{equation}
where $E(m)$ is the elliptic integral of second kind~\cite{abra}. Eq.~(\ref{e1}) is just the well known result for the frequency of a pendulum, while Eq.~(\ref{e2}) expresses the fact that the particle's action is $I_f$~\cite{benisti07}.    

To conclude the derivation of $\langle \sin(x) \rangle$, we use Liouville theorem, $\tilde{f}(\theta,I,t)= \tilde{f}[\theta_M(\theta,I),I_M(\theta,I),t_M]$, we approximate the distribution in action by a delta function at $I=I_f$, and we expand $\tilde{f}(\theta_M,I_M)$ in Fourier series to find,
\begin{eqnarray}
\nonumber
\tilde{f}(\theta,I,t)&=& \tilde{f}[\theta_M(\theta,I),I_M(\theta,I),t_M]\\
\label{e3}
&=& \sum_{n=0}^{+\infty} \left[Êf_{cn} \cos(n\theta_M) +f_{sn}Ê\sin(n\theta_M) \right] \delta(I_M-I_f).
\end{eqnarray}
Plugging Eqs.~(\ref{e0})~and~(\ref{e3}) into the expression~(\ref{0})~for~$\langle \sin(x) \rangle$, taking advantage of the fact that the Jacobian of the change of variables $(\theta,I) \rightarrow (\theta_M,I_M)$ is unity, and using the value of $\langle \sin(x) \rangle$ at $t=t_M$ when $\Phi=\Phi_M$ derived from perturbation theory, namely $\langle \sin(x) \rangle=S_M$ when $t=t_M$, we find,
\begin{equation}
\label{res1}
\langle \sin(x) \rangle=S_M\frac{K^2_M}{K^2}Ê\sqrt{\frac{q}{q_M}}\frac{1-q_M}{1-q} \cos\left[\int_{t_M}^t \omega_0(I_f)dt'Ê\right],
\end{equation}
where $K_M$ and $q_M$ are the values of $K$ and $q$ when $\Phi=\Phi_M$. Note that, in Eq.~(\ref{res1}), we did not account for the term proportional to $\sin \left[\int_{t_M}^t \omega_0(I_f)dt'Ê\right]$. This term should actually be negligible because the value reached by $\langle \sin(x) \rangle$ at $t=t_M$ is a local maximum, and $\int_{t_M}^t \omega_0(I_f)dt'$ varies much more rapidly with time than $q$. Therefore, the maxima of $\langle \sin(x) \rangle$ are identified with those of $\cos\left[\int_{t_M}^t \omega_0(I_f)dt'Ê\right]$. 

We now make the change of variables $t\rightarrow \Phi$ in the integral of Eq.~(\ref{res1}) to find, when $\Phi$ grows exponentially in time, 
\begin{equation}
\label{res2}
\langle \sin(x) \rangle \approx S_M\frac{K^2_M}{K^2}Ê\sqrt{\frac{q}{q_M}}\frac{1-q_M}{1-q} \cos\left[\int_{\Phi_M}^\Phi \frac{d\Phi'}{\sqrt{\Phi'}K[m(I_f)]}Ê\right],
\end{equation}
where $m(I_f)$ is related to $\Phi$ by Eq.~(\ref{e2}). 

\begin{figure}[!h]
\centerline{\includegraphics[width=12cm]{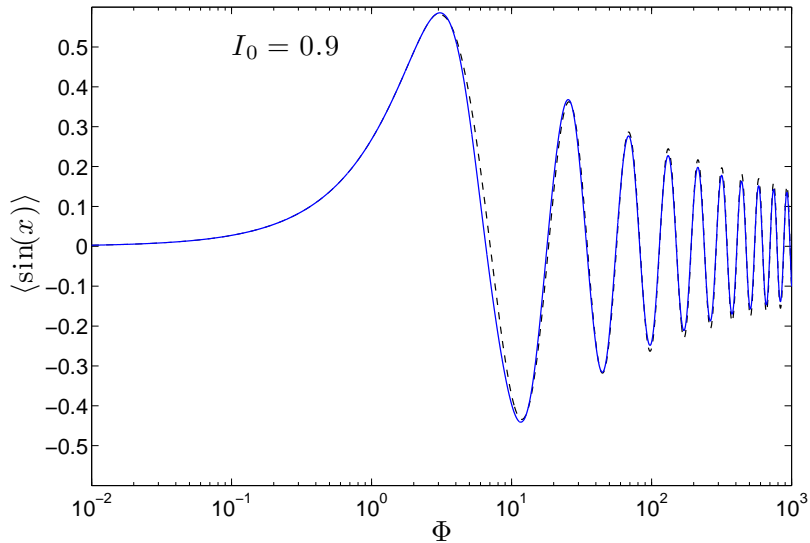}}
\caption{\label{f8} $\langle \sin(x) \rangle$ when $I_0=0.9$ as calculated numerically (blue solid line) and theoretically (black dashed line) by connecting the pertubative estimate with the values of $\langle \sin(x) \rangle$ given by Eq.~(\ref{res2}).}
\end{figure}

As shown in Fig.~\ref{f8}, the values of $\langle \sin(x) \rangle$ for $I_0=0.9$ obtained by using the perturbative estimate of Ref.~\cite{benisti07} for $\Phi \leq \Phi_M \approx 3.08$, and Eq.~(\ref{res2}) for $\Phi \geq \Phi_M$, agree very well with the numerical ones. In order to derive the value for $I_f$, we followed the method described in Section~\ref{II}, i.e., using a perturbation analysis, we estimated the most probable action when $\Phi=\Phi_M$. When doing so, we found $I_fÊ\approx 0.495$, so that the global change in action, $\Delta I$, is about 40\% of $I_0$.  

\subsubsection{Use of $\chi_i$ as a diagnostic for $\Delta I$}

\begin{figure}[!h]
\centerline{\includegraphics[width=12cm]{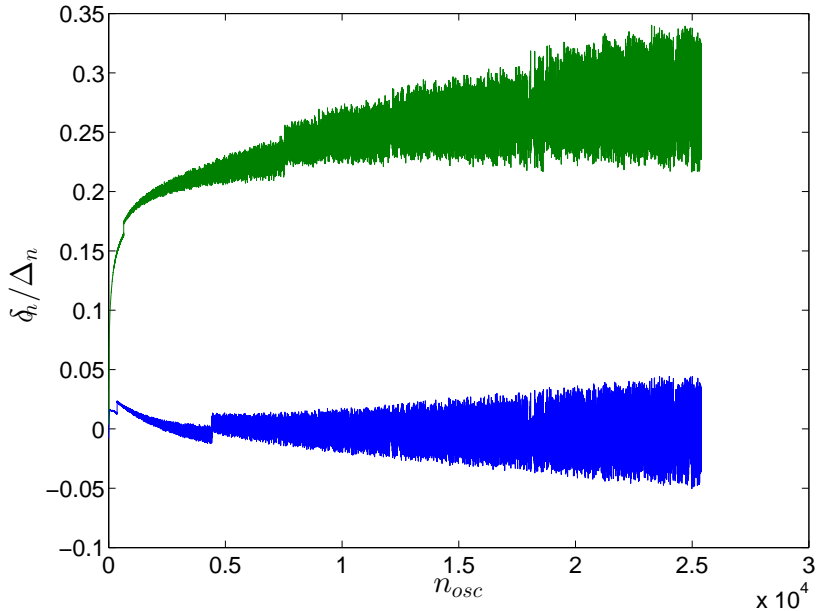}}
\caption{\label{f9} $\delta_n/\Delta_n$ as a function of the number of oscillations, $n_{osc}$, when $I_0=100$, and by using in Eq.~(\ref{res2}) $I_f=99.56$ (blue line) or $I_f=100$ (green line).}
\end{figure}

\begin{figure}[!h]
\centerline{\includegraphics[width=12cm]{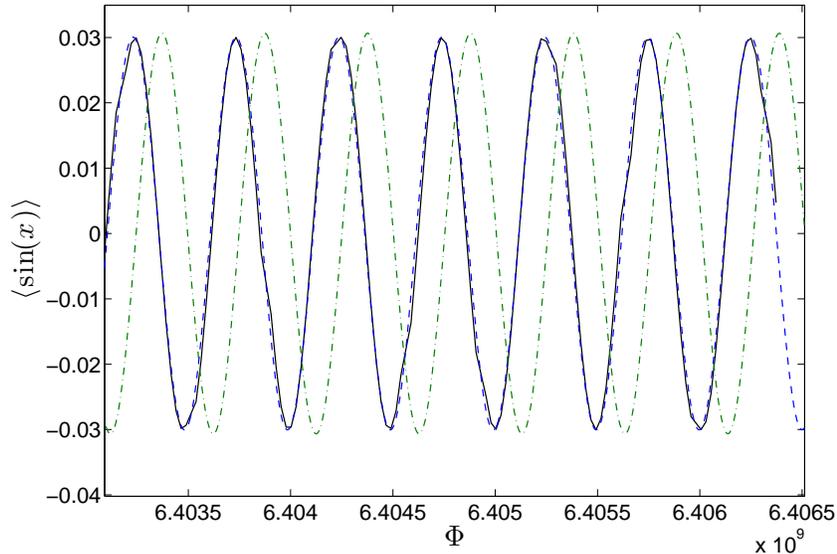}}
\caption{\label{f10} Comparisons between the values of $\langle \sin(x) \rangle$ when $I_0=100$ calculated numerically (black solid line) and by making use of Eq.~(\ref{res2}) with $I_f=99.56$ (blue dashed line) and $I_f=100$ (green dash-dotted line) after 23,595 oscillations.}
\end{figure}
As shown in Fig.~\ref{f8}, $\langle \sin(x) \rangle$ oscillates very quickly with $\Phi$ so that a small error in the estimate of the frequency, $\omega_0$, of these oscillations would entail a shift in the positions of the maxima of $\langle \sin(x) \rangle$ that would be rapidly visible. Therefore, one way to guess $I_f$ numerically may consist in trying to match the locations of the maxima of  $\langle \sin(x) \rangle$, obtained from numerical simulations, with those derived from Eq.~(\ref{res2}). More precisely, we denote respectively by $\Phi_n^{num}$ and $\Phi_n^{th}$ the numerical and theoretical estimates of the amplitude at which  $\langle \sin(x) \rangle$ reaches its $n^{th}$ maximum, and we introduce $\Delta_n \equiv \Phi_{n+1}^{num}-\Phi_n^{num}$ and $\delta_n=\Phi_n^{th}-\Phi_n^{num}$. Then, $I_f$ is found numerically as the value which, when used in Eq.~(\ref{res2}), makes the ratio $\delta_n/\Delta_n$ as small as possible over a large number of oscillations.  It is actually very demanding to calculate $\langle \sin(x) \rangle$ up to very large amplitudes, and, to do so, we had to use a very small time step, $dt=10^{-9}$. Fig.~\ref{f9} plots $\delta_n/\Delta_n$ when $I_0=100$ and $I_f=99.56$ (blue line) or $I_f=100$ (green line). It shows that, when using $I_f=99.56$ in Eq.~(\ref{res2}) (with $\Phi_M$ and $S_M$ obtained numerically), the positions of the maxima of $\langle \sin(x) \rangle$ remain very close to the numerical ones. Indeed, even after the 25,395 oscillations we calculated, they differ by less than the uncertainty due to the discreteness of $\langle \sin(x) \rangle$, and the averaged value $\delta_n/\Delta_n$ is found to be close to $-8\times10^{-4}$. The good agreement between the values of  $\Phi_n^{num}$ and  $\Phi_n^{th}$ with  $I_f=99.56$ may also be appreciated in Fig.~\ref{f10} [where we multiplied the amplitudes of $\langle \sin(x) \rangle$ given by Eq.~(\ref{res2}) by a factor close to 0.8, because they were slightly overestimated by this equation  for the large amplitudes considered in Fig.~\ref{f10}, as discussed in the end of Appendix~\ref{A}]. Moreover, as shown in Figs.~\ref{f9}~and~\ref{f10}, neglecting the change in action and using $I_f=I_0=100$ instead of $I_f=99.56$ does entail a shift in the locations of the maxima of $\langle \sin(x) \rangle$ that is clearly visible. Hence, in addition to being an important physics quantity, $\chi_i$ may be used as a very fine diagnostic that reveals a relative error in the particles global action as small as 0.5\%. Moreover, the results of Figs.~\ref{f8}-\ref{f10} clearly show that the very concept of a global action for a set of particles is relevant to theoretically compute macroscopic quantities such as $\chi_i$, and, in particular, the contribution to $\chi_i$ from trapped particles. Actually, for the parameters of Fig.~\ref{f9}, making use of Eq.~(\ref{res2}) reduces the computation time of $\langle \sin(x) \rangle$ by more than 4 orders of magnitude compared to a direct numerical resolution of the equations of motion.

\subsection{Use of $\chi_i$ to compute the global change in action}
In the previous Subsection we saw that we could compute $\chi_i$ very efficiently by making use of the concept of global action, provided that $\Delta I$ was known. In Section~\ref{II},  $\Delta I$ was obtained theoretically, for small enough values of $I_0$, from the particles' distribution function derived by making use of a perturbation analysis. Now, it is much more difficult to derive the distribution function than to estimate one of its Fourier coefficient, so that the perturbative values of $\langle \sin(x) \rangle$, and in particular the estimates of $\Phi_M$ and $S_M$, are expected to be accurate for a larger range in $I_0$ than the distribution function itself. Moreover, from the perturbative estimate of  $\langle \sin(x) \rangle$ it is possible calculate $\Delta I$, as we shall now show it.

Plugging Eqs.~(\ref{e0}), (\ref{ee0}) and (\ref{e3}) into Eq.~(\ref{0}) yields,
\begin{equation}
\label{C1}
\langle \sin(x) \rangle=\pi^2 f_{s1} \frac{\sqrt{q}}{K^2(1-q)} \cos\left[\int_{t_M}^t \omega_0(I_f)dt'Ê\right],
\end{equation}
showing that the maxima of $\langle \sin(x) \rangle$ are proportional to $\sqrt{q}/K^2(1-q)$. Moreover, as discussed in the Appendix~\ref{A}, the coefficient $ f_{s1}$ depends very little on $I_0$, and may therefore be considered as a constant. This is illustrated in Fig.~\ref{f11} showing that the local maxima of $\langle \sin(x) \rangle$, when plotted as a function of $m$, lie on a curve that depends very little on $I_0$, and that is close to that given by Eq.~(\ref{C2}) with $S_0=11.6$ (see the end of Appendix~\ref{A} for a discussion of this value). 

\begin{figure}[!h]
\centerline{\includegraphics[width=19cm]{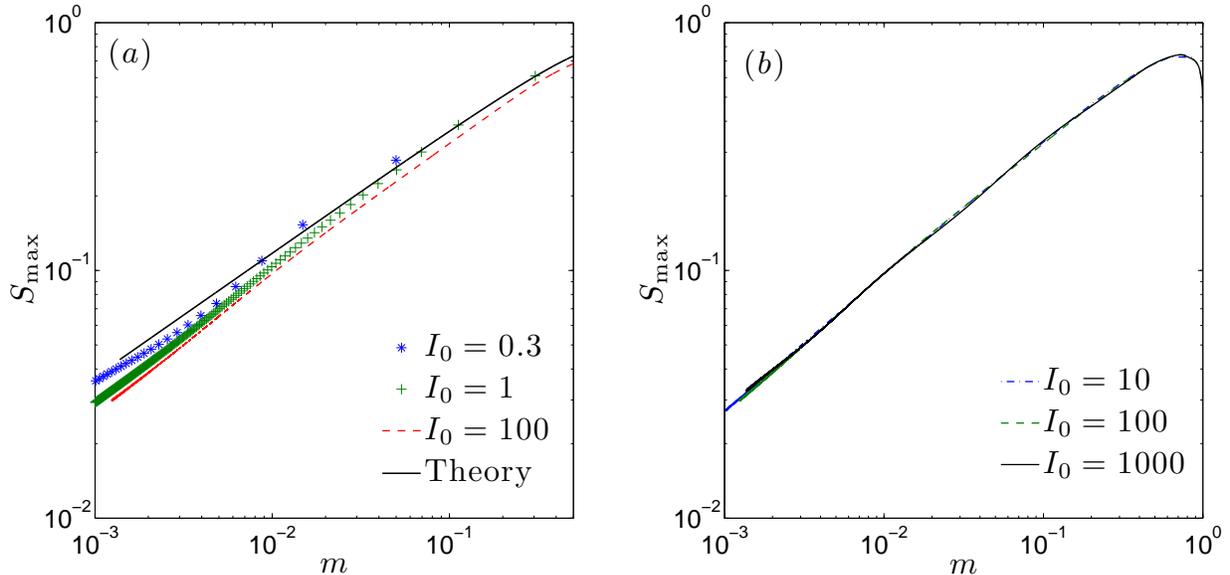}}
\caption{\label{f11}Local maxima, $S_{\max}$, of $\langle \sin(x) \rangle$ as a function of $m$ for various values of $I_0$. The black solid curve in panel (a) plots the values given by Eq.~(\ref{C2}) with $S_0=11.6$.}
\end{figure} 

Using this result, we conclude that the first maximum, $S_M$, of $\langle \sin(x) \rangle$ is such that,
\begin{equation}
\label{C2}
S_M=S_0\frac{\sqrt{q_M}}{K_M^2(1-q_M)},
\end{equation}
where $S_0$ is a constant, and where $q_M \equiv q(m_M)$ and $K_M \equiv K(m_M)$, with $m_M$ such that,
\begin{equation}
\label{C3}
\frac{4}{\pi}Ê\sqrt{\Phi_M}Ê\left[ÊE(m_M)+(m_M-1)K(m_M)Ê\right]=I_f.
\end{equation}
Using a perturbation analysis, we are able to estimate $S_M$ and $\Phi_M$. Therefore, if the constant $S_0$ is known, one just has to solve Eq.~(\ref{C2}) for $m_M$, and to plug the value thus found in Eq.~(\ref{C3}) in order to calculate $I_f$. In order to derive the constant $S_0$, we need to know the change in action, $\Delta I= I_f-I_0$, at least for one $I_0$, which we do by making use of the method described in Sec.~\ref{II}, and which yields $S_0 \approx 11.6$ [the accuracy of this estimate may be appreciated in Fig.~\ref{f11}~(a)]. With this value of $S_0$, we calculate the function $\Delta I (I_0)$ represented by the dashed curve in Fig.~\ref{f6} (a), and which appears to be accurate when $I_0 \alt 1.6$. To be more specific, from Eqs.~(\ref{C2}) and~(\ref{C3}) with $\Phi_M$ and $S_M$ estimated perturbatively, we find that, when $I_0=1.6$, the change in action should be $\Delta I  \approx -0.444$. This is to be compared with the numerical result $\Delta I  \approx -0.437$ given at the end of Section~\ref{II}, and with the neo-adiabatic estimate, $\Delta I  =-(2/\pi)\ln(2)\approx -0.441$. We therefore conclude that, by making use of a perturbation analysis, it is possible to provide accurate estimates of the global change in action up to values of $I_0$ large enough for the neo-adiabatic estimate to be also very accurate.  This shows that, indeed, the action change due to trapping can be calculated theoretically, whatever the range of variation of the dynamics, by connecting the perturbative results with the neo-adiabatic ones. 

\section{Conclusion}
\label{iv}
In this paper, we introduced the concept of a ``global action'' for a set of particles with same initial action, $I_0$. This was done by showing that, when all the particles are trapped, the distribution in action, $f(I)$, has one very sharp peak, at the smallest action. In addition to numerical evidences, this result was proved theoretically by making use of a perturbation analysis in the potential amplitude, which is valid when $I_0 \alt 1$, and by making use of the neo-adiabatic theory, which is already quite accurate when $I_0 \agt 1$. Moreover, we showed that the global action we defined was relevant, and actually very useful, to efficiently compute macroscopic quantities, such as the imaginary part, $\chi_i$, of the electron susceptibility for a plasma wave. In particular, we could compute very accurately $\chi_i$ whether the particles were trapped or untrapped, even when the particles' motion was far from adiabatic before trapping, a result that was not available in previous publications~\cite{benisti07,yampo,dodin2}. As for the change in action, $\Delta I =(I_f-I_0)$, we could derive it whatever the rate of variation of the dynamics and, in particular, for a non slowly varying dynamics, when $\Delta I$ was not small compared to $I_0$.  To the best of our knowledge, no equivalent result has ever been published since the change of action due to trapping has always been estimated by making use of the neo-adiabatic theory, that is only accurate for slowly varying dynamics. Our derivation of $\Delta I$ mainly rests on a pertubative expansion, in the potential amplitude, of the particles' motion. More precisely, it is derived by plugging the perturbative estimate of the first maximum, $S_M$, of $\langle \sin(x) \rangle$ (which is proportional to $\chi_i$), and of the corresponding value of the potential amplitude, $\Phi_M$, into Eqs.~(\ref{C2}) and (\ref{C3}) with $S_0 \approx 11.6$. These equations provide an accurate estimate of $\Delta I$ up to the point when it becomes essentially independent of $I_0$, and nearly matches the constant value $\Delta I \approx -(2/\pi)\ln(2)$ provided by the neoadiabatic theory. The latter value was, moreover, found to be in excellent agreement with numerical results. 

In conclusion, this paper shows two main results which, we believe, are completely new. (i) the notion of a global action and its relevance to theoretically compute macroscopic quantities such as $\chi_i$ ; (ii) the theoretical derivation of the global change in action due to trapping, whether the dynamics is slowly varying or not. Moreover, as shall be shown in a forthcoming paper, the results derived here, in particular as regards the theoretical estimate of $\chi_i$, constitute an essential step to describe the nonlinear regime of the beam-plasma instability and to theoretically compute the nonlinear Landau damping rate of a plasma wave, which are long standing issues in plasma physics. 

\begin{acknowledgments}
 One of the authors (D.B) would like to thank D.F. Escande for a careful reading of the manuscript, for very useful comments, and for pointing out the results of Ref.~\cite{nonlinearity}. 
 \end{acknowledgments} 

\appendix
\section{Shift in angle entailed by trapping}
\label{A}
\setcounter{equation}{0}
\newcounter{app}
\setcounter{app}{1}

\setcounter{app}{1}
\label{A}
In this Appendix, we show that the action change due to trapping entails  a shift in the variation of the angle, compared to a purely adiabatic motion, which is of the order of unity, and is essentially independent of $I_0$ in the limit $I_0 \rightarrow \infty$. This proves that, once the action has converged towards a nearly constant value, the distribution in angle, which would have been uniform for a purely adiabatic motion, changes in a fashion that is essentially independent of $I_0$. Consequently, the Fourier component of this distribution, which we denoted by $f_{s1}$ in Sec.~\ref{III}, is essentially independent of $I_0$ so that the factor $S_0$ in Eq.~(\ref{C2}) for $\langle \sin(x) \rangle$ is indeed a constant, as illustrated in Fig.~({\ref{f11}) of Section~\ref{III}.

In order to show the aforementioned results, we calculate the variation in angle up to a time, $t_t$, when the shift in action has reached its asymptotic value, $\Delta I$, and the particles are deeply trapped [since Eq.~(\ref{C2}) on Section~\ref{III} is only valid in this limit]. Using the well know results that, for an untrapped particle, 
\begin{equation}
\frac{d\theta}{dt}=\frac{\pi \sqrt{\Phi}}{2\sqrt{m}K(m)},Ê
\end{equation}
where $m$ is related to the action and the amplitude by
\begin{equation}
\label{B1}
\frac{4\sqrt{\Phi}}{\pi \sqrt{m}}ÊE(m)=I,
\end{equation}
while, for a trapped particle, 
\begin{equation}
\frac{d\theta}{dt}=\frac{\pi \sqrt{\Phi}}{2K(m)},Ê
\end{equation}
with
\begin{equation}
\label{A0}
\frac{4\sqrt{\Phi}}{\pi}\left[ÊE(m)+(m-1)K(m) \right] =I,
\end{equation}
one easily finds (using $d\Phi/dt=\Phi$), 
\begin{equation}
\label{dtheta}
\Delta \theta=\frac{\pi^2}{8}\left\{\int_{m_{\min}}^1\frac{I}{mE^2}dm+\int_{m_t}^1\frac{I}{\left[E+(m-1)K\right]^2}dm\right\},
\end{equation}
where the first term accounts for the variation of the angle while the particle is untrapped, and the second term is the angle variation when the particle is trapped. In Eq.~(\ref{dtheta}), $m_{\min}$ is the value of $m$ defined by Eq.~(\ref{B1}) at $t=0$ when $\Phi=\Phi_0$ and $I=I_0$, while $m_t$ is the value of $m$ defined by Eq.~(\ref{A0}) at time $t_t$ when $I\approx I_0-\Delta I$. Then, from Eq.~(\ref{dtheta}), it is easily found that the action change entails the following shift in $\Delta \theta$,
\begin{equation}
\label{ddtheta}
\delta(\Delta \theta)=\frac{\pi^2}{8}\left\{\int_{m_{\min}}^1\delta \left[\frac{I}{mE^2}\right]dm+\int_{m_t}^1\delta \left[\frac{I}{\left[E+(m-1)K\right]^2}\right]dm-\frac{I\delta m_t}{\left[E+(m_t-1)K\right]^2}\right\}.
\end{equation}
Let us now denote by $\delta I_m$ the instantaneous change in action, when the wave amplitude assumes the value $\Phi$, i.e., $\delta I_m \equiv I[m(\Phi)]-I_0$. From Eq.~(\ref{B1}), it is easily found that, when the particle is untapped, the change $\delta I_m$ entails a change in $m$ by,
\begin{equation}
\label{A1}
\delta m=-\frac{2mE}{K}\frac{\delta I_m}{I},
\end{equation}
while, when the particle is trapped, from~Eq.~(\ref{A0}), one finds,
\begin{equation}
\label{A2}
\delta m =\frac{2[E+(m-1)K]}{K}\frac{\delta I_m}{I}.
\end{equation}
Plugging the results from Eqs.~(\ref{A1}) and (\ref{A2}) into Eq.~(\ref{ddtheta}), one easily finds, 
\begin{equation}
\frac{8\delta(\Delta \theta)}{\pi^2}=\int_{m_{\min}}^1\frac{\delta I_m}{mE^2}\left[\frac{4E}{K}-1\right]dm-\int_{m_t}^1\frac{\delta I_m}{\left[E+(m-1)K\right]^2}dm -\frac{2\delta I_{m_t}}{K[E+(m_t-1)K]}.
\end{equation}
When the particle is untrapped, $\delta I_m$ is negligible except close to the separatrix, where it scales as $\ln(I_0)$ when $(1-m) \sim 1/I_0$. Similarly, when the particle is trapped then, within a narrow region close to the separatrix where $(1-m)\sim 1/I_0$, $\delta I_m$ scales as $\ln(I_0)$. However, away from this narrow region, $\delta I_m$ for a trapped particle is very close to its asymptotic value, $\Delta I$. Hence, one finds,
\begin{equation}
\label{A3}
\delta(\Delta \theta)=-\frac{\pi^2Ê\Delta I }{8}F(m_t)+O\left[\frac{\ln(I_0)}{I_0}\right],
\end{equation}
where 
\begin{equation}
\label{A4}
F(m_t)=\int_{m_t}^1\frac{\delta I_m}{\left[E+(m-1)K\right]^2}dm +\frac{2\delta I_{m_t}}{K[E+(m_t-1)K]}.
\end{equation}
For large $I_0$'s, since $\Delta I$ becomes essentially independent of $I_0$, it is clear form Eq.~(\ref{A3}) that so does $\delta(\Delta \theta)$. The entails that the values of the Fourier coefficient, $f_{s1}$, as a function of $m$, should become independent of $I_0$ as $I_0$ increases, in agreement with the results of Fig.~({\ref{f11}) of Section~\ref{III}. 

Note that $f_{s1}$ is necessarily less than unity, while we used $\pi^2f_{s1} \equiv S_0 \approx 11.6$ in Eq.~(\ref{C2}). This value was used in order to derive precisely the amplitude of the first maximum of $\langle \sin(x) \rangle$, which occurs for an amplitude which is not large enough for the approximate expression of $\sin(x)$ given by Eq.~(\ref{e0}) of Section~\ref{III} to be extremely accurate (although it is already a good approximation). For subsequent maxima, which occur for large amplitudes and, therefore, small values for $m$, the expression for $\langle \sin(x) \rangle$ given by Eq.~(\ref{C1}) becomes very accurate, and, in this expression $f_{s1}$ is necessarily a constant (independent of $m$) less than unity. This explains why, in Fig.~\ref{f10} of Section~\ref{III}, we had to multiply the amplitude of $\langle \sin(x) \rangle$, as given by Eq.~(\ref{res2}), by a constant close to 0.8 in order to match the numerical results (and one may actually notice that $\pi^2/11.6 \approx 0.85$).


\begin{thebibliography}{10}
\bibitem{explication}
There was, a priori, no reason to stop at order 12, except the memory limitations we had to face when running our routine for symbolic computation. However, going to order 12 was enough for the purpose of this paper, and, therefore, we did not try to go beyond this order. As regards the computation of $\langle \sin(x) \rangle$, only odd orders give a nonzero contribution so that its value at order 11 is the same as that obtained at order 12.
\bibitem{explication2}
When plotting the perturbative estimate of $\langle \sin(x) \rangle$ as a function of $\Phi$, we actually only find a local maximum when $I_0\alt 1.2$, which is a defect of the perturbative expansion. When $I_0\agt1.2$, the first maximum of $\langle \sin(x) \rangle$, and the amplitude when it is reached, are estimated by the point where the slope of $\langle \sin(x) \rangle(\Phi)$, as calculated perturbatively, is minimum. This estimate remains accurate as long as $I_0 \alt 1.6$, and, for these values of $I_0$, it does provide a good estimate of the jump in action, as shown in Section~\ref{III}.
\end{thebibliography}
\end{document}